\documentclass[preprint,authoryear,11pt,5p,times]{elsarticle}
\usepackage{times}
\usepackage{bm}
\usepackage{amsmath}
\usepackage{latexsym}%
\usepackage{graphicx}%
\usepackage{amssymb}%
\usepackage{graphicx}
\usepackage{booktabs}
\usepackage[breaklinks=true,colorlinks=true]{hyperref}
\usepackage{listings}
\usepackage{algorithm}
\usepackage{algpseudocode}
\usepackage{url}

\usepackage[multiple]{footmisc}
\usepackage[table]{xcolor}
\usepackage[many]{tcolorbox}
\usepackage{listings}
\definecolor{ballblue}{rgb}{0.8, 0.25, 0.33}
\definecolor{tealblue}{rgb}{0.21, 0.46, 0.53}

\definecolor{codegreen}{rgb}{0,0.6,0}
\definecolor{codegray}{rgb}{0.5,0.5,0.5}
\definecolor{codepurple}{rgb}{0.58,0,0.82}
\definecolor{backcolour}{rgb}{0.95,0.95,0.92}
\lstdefinestyle{mystyle}{
  backgroundcolor=\color{backcolour},   commentstyle=\color{codegreen},
  keywordstyle=\color{magenta},
  numberstyle=\tiny\color{codegray},
  stringstyle=\color{codepurple},
  basicstyle=\ttfamily\footnotesize,
  breakatwhitespace=false,         
  breaklines=true,                 
  captionpos=b,                    
  keepspaces=true,                 
  numbers=left,                    
  numbersep=5pt,                  
  showspaces=false,                
  showstringspaces=false,
  showtabs=false,                  
  tabsize=2
}

\journal{Astronomy And Computing}

\begin{document}

\begin{frontmatter}

\title{\texttt{qrpca:} A Package for Fast Principal Component Analysis with GPU Acceleration}

\author[1]{Rafael S. de Souza\corref{mycorrespondingauthor}}
\cortext[mycorrespondingauthor]{Corresponding author}
\ead{drsouza@shao.ac.cn}

\author[1]{Xu Quanfeng 
}
\author[1,2]{Shiyin Shen 
}
\author[1,3]{Chen Peng 
}
\author[1]{Zihao Mu 
}

\affiliation[1]{Key Laboratory for Research in Galaxies and Cosmology, Shanghai Astronomical Observatory, Chinese Academy of Sciences, 80 Nandan Rd., Shanghai 200030, China
}
\affiliation[2]{Key Lab for Astrophysics, Shanghai, 200034, People's Republic of China}
\affiliation[3]{Shanghai Institute of Technology, 100 Haiquan Rd., Shanghai 201418, China
}

\begin{abstract}
We present \texttt{qrpca}, a fast and scalable  QR-decomposition principal component analysis package.
The software, written in both \texttt{R} and \texttt{python} languages, makes use of  \texttt{torch} for internal matrix computations, and enables \texttt{GPU} acceleration, when available.  \texttt{qrpca} provides similar functionalities to \texttt{prcomp} (\texttt{R}) and \texttt{sklearn} (\texttt{python}) packages respectively. A benchmark test shows that \texttt{qrpca} can achieve computational speeds 10-20 $\times$ faster for large dimensional matrices than default implementations, and is at least twice as fast for a standard decomposition of spectral data cubes. The \texttt{qrpca} source code is made freely available to the community. 
\end{abstract}

\begin{keyword}
Principal component analysis; Astroinformatics; GPU computing    
\end{keyword}

\end{frontmatter}

\section{Introduction}
\label{sec:introduction}

Principal component analysis \citep[PCA;][]{Pearson1901} stands out as a prime method for dimensionality reduction and data exploration \citep[see][for a review]{Jollife2016}. It compresses a dataset while preserving as much variability as possible. Given the original matrix input, PCA performs either eigendecomposition or singular value decomposition (SVD) and outputs a matrix comprising orthogonal variables, which are linear combinations of the original variables,  termed principal components (PCs). 

The technique is among the most popular tools in Astronomy and has been applied to a broad range of studies. Notable examples include the analysis of exoplanets imaging data \citep{Amara2012,Kiefer2021}, Type Ia supernova photometric classification \citep{Ishida2013},  analysis of cosmological simulations \citep{DeSouza2014},  foreground separation of 21 cm intensity maps  \citep{Yohana2021}, and point spread function reconstruction \citep{Nie2021}.

Despite its broad applicability, SVD PCA  implementations are computationally costly for high dimensional matrices\footnote{Throughout the paper, we will refer to data size as the number of rows, and dimension as the number of columns.}. This limitation triggered the development of PCA extensions, including sparsity assumptions \citep[e.g.][]{Fan2018,ADNAN2021101710}, and massive parallelism \citep{LAZCANO2017101}. \citet{HashTreePC} suggested a hash-tree PCA to accelerate conventional PCA by sampling similar objects while preserving the original data distribution. \citet{VOGT20011} suggested a two-step procedure for analyzing hyperspectral images, where the algorithm first compresses the spectra via wavelets transform before applying a conventional SVD. 

This paper contributes to the literature in two main aspects. To the best of our knowledge, we present the first public package for  QR-decomposition PCA, and the package allows seamless {\tt GPU} acceleration. $\mathbf{QR}$ factorization is a decomposition of a matrix $\mathbf{A}$ into a product $\mathbf{A} = \mathbf{QR}$ of an orthogonal matrix $\mathbf{Q}$ and an upper triangular matrix $\mathbf{R}$.

The user-friendly implementation named  \texttt{qrpca} uses \texttt{torch} \citep{RN7,rtorch} under the hood for matrix computations. The package is particularly suited for large dimensional matrices and provides similar functionalities to other major distributions such as \texttt{prcomp} in \texttt{R}, and \texttt{sklearn} in \texttt{python}. 

The remainder of the paper is organized as follows. In \autoref{sec:methods} we provide an overview of the SVD PCA and  QR-decomposition PCA. \autoref{sec:results} shows a speed performance test alongside a practical application example to integral field unit (IFU) spectroscopy. Finally, in \autoref{sec:conclusion} we present our concluding remarks.

\section{Methodology}
\label{sec:methods}

Let $\mathbf{X} \in \mathbb{R}^{n\times m}$ be a rectangular matrix composed by  $n$ rows by  $m$ columns. The SVD decomposition of  $\mathbf{X}$ is given by:
\begin{equation}
\mathbf{X} =  \mathbf{U}\mathbf{S}\mathbf{V}^{\intercal}. 
\end{equation}
Here, $\mathbf{S} \in \mathbb{R}^{m\times m}$ is a diagonal matrix, and  $\Phi =  \mathbf{U}\mathbf{S}$ gives the PCA projection.   We show at Algorithm  \ref{alg:svdca} a pseudo-code for a SVD PCA.  If the dimensionality of $\mathbf{X}$  is large, then the computation of the eigenvalues will be time consuming, and may cause memory overflow \citep[e.g.][]{ADNAN2021101710}.

A faster alternative is to use an intermediate step as suggested by \citet{RN1}. The procedure consists in first factorize $\mathbf{X}$ into an  orthogonal matrix $\mathbf{Q} \in \mathbb{R}^{n\times m}$, and an upper triangular matrix $\mathbf{R} \in \mathbb{R}^{m\times m}$. The SVD decomposition of  $\mathbf{R}^{\intercal} = \mathbf{U}_{\star}\mathbf{S}_{\star}\mathbf{V}^{\intercal}_{\star}$ then yields the same diagonal matrix $\mathbf{S}_{\star} \equiv  \mathbf{S} $ of $\mathbf{X}$, and an equivalent PCA transform $\mathbf{Q}\mathbf{V}\mathbf{S} \equiv \mathbf{U}\mathbf{S}$. 
The computational advantage comes from the intermediate  $\mathbf{Q}\mathbf{R}$ decomposition, which enables running SVD factorization on the upper triangular matrix $\mathbf{R}$ to compute eigenvectors instead of running SVD directly on $\mathbf{X}$. See Algorithm  \ref{alg:qrpca} for a  pseudo-code for the case of  QR-decomposition PCA.

\begin{algorithm}[ht]

    \caption{SVD PCA}\label{alg:svdca}
    \hspace*{\algorithmicindent} 
    \begin{algorithmic}[1]
        \Require Input matrix $\mathbf{X} \in \mathbb{R}^{n\times m}$
    \State Compute SVD on $\mathbf{X}$\\
    $\mathbf{U}$ $\in$ $\mathbb{R}^{n\times m}$\quad Orthogonal matrix  \\
       $\mathbf{S}$ $\in$ $\mathbb{R}^{m\times m}$ \quad Rectangular diagonal matrix  \\
        $\mathbf{V}$ $\in$ $\mathbb{R}^{m\times m}$ \quad Orthogonal matrix
    \State Compute PCA transform $\Phi$ = $\mathbf{U}\mathbf{S}$ 
    \end{algorithmic}
\end{algorithm}

\begin{algorithm}[ht]
  \caption{QR-decomposition PCA}\label{alg:qrpca}
    \hspace*{\algorithmicindent} 
    \begin{algorithmic}[1]
        \Require Input matrix $\mathbf{X} \in \mathbb{R}^{n\times m}$
        \State Compute $\mathbf{QR}$ decomposition on $\mathbf{X}$\\
        $\mathbf{Q}$ $\in$ $\mathbb{R}^{n\times m}$ \quad Orthogonal matrix \\
        $\mathbf{R}$ $\in$ $\mathbb{R}^{m\times m}$ \quad Upper triangular matrix
    \State Compute SVD on $\mathbf{R}^{\intercal} \in \mathbb{R}^{m\times m}$\\
       $\mathbf{S}$ $\in$ $\mathbb{R}^{m\times m}$  \\
        $\mathbf{V}$ $\in$ $\mathbb{R}^{m\times m}$ 
    \State Compute PCA transform $\Phi$ = $\mathbf{Q}\mathbf{V}\mathbf{S}$ 
    \end{algorithmic}
\end{algorithm}

\section{Performance evaluation}
\label{sec:results}
In this section, we provide a simple test of the  \texttt{qrpca} computational performance on simulated data and an example application to the analysis of Astronomical IFU spectra.

\subsection{Simulated data}

\begin{figure*}[ht]
 \centering
\includegraphics[width=0.95\linewidth]{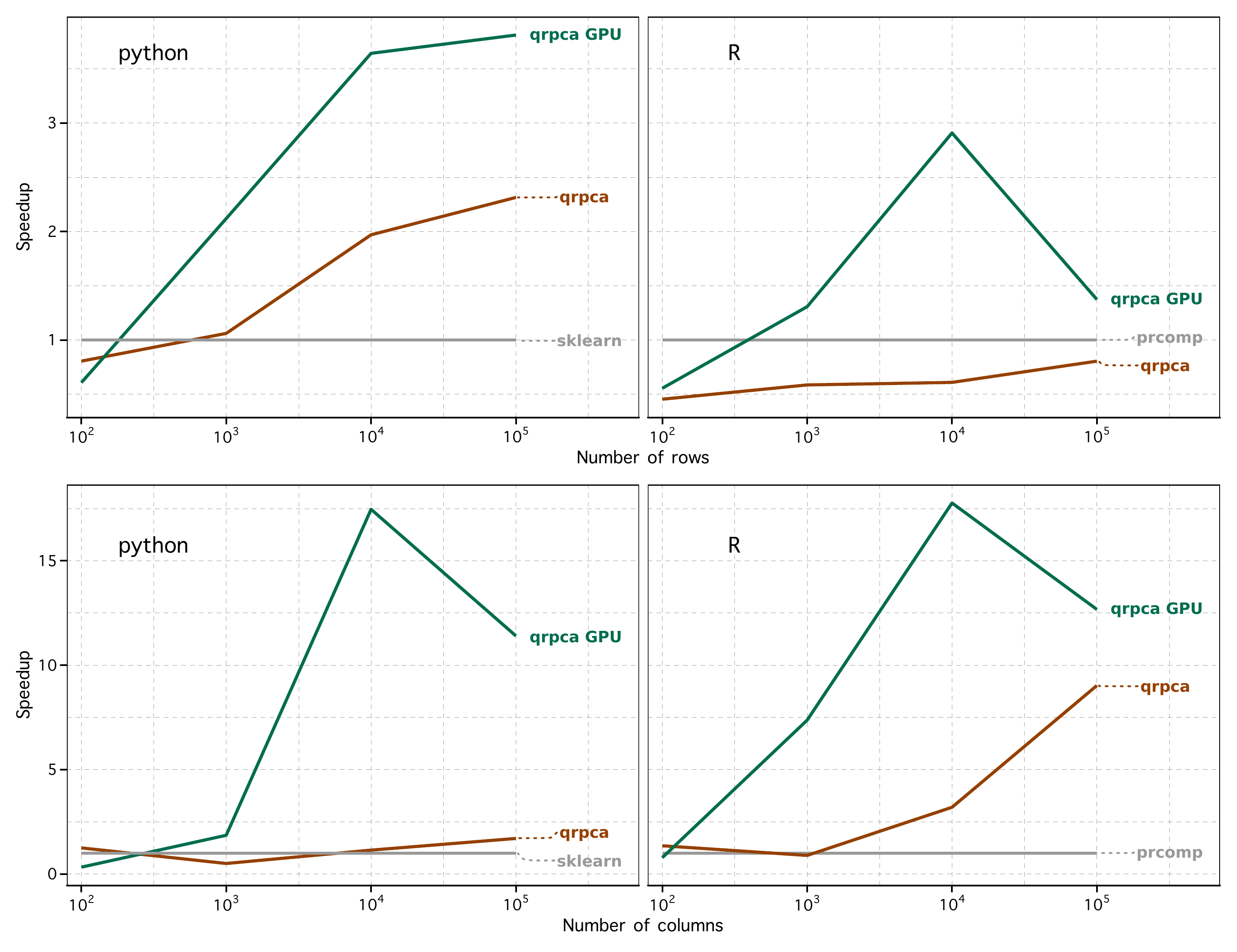}
\caption{A speedup performance evaluation between \texttt{qrpca} and  major SVD PCA implementations, namely {\tt sklearn} and {\tt prcomp}. Each line represents the speedup time normalized by the running time of {\tt sklearn} (left panels), and {\tt prcomp} (right panels). 
The benchmark was executed in a machine with the following specifications: CPU -- 2.2GHz Intel Xeon Silver 4210; GPU -- NVIDIA A40 (48GB); OS --  Ubuntu Linux 18.04 64 bits; RAM -- 188 GB. }
\label{fig:speed}
\end{figure*}

To evaluate the performance of \texttt{qrpca}, we create a collection of random matrices varying the rows and column sizes. For a fixed dimension of $m = 1000$, we vary the data size from $n = 10^2 - 10^6$, and for a fixed data size of $n = 1,000$, we varied the dimensions from $m = 10^2 - 10^5$. The mock data is sampled from a normal zero mean and unity variance distribution. Fig. \ref{fig:speed} shows the speedup gain for a range of dataset sizes and dimensions. Left panels show the comparison between the python version of \texttt{qrpca} and default PCA implementation of  {\tt sklearn}, while the right panels show the comparison between the R implementation of \texttt{qrpca} and \texttt{prcomp}.  Visual inspection on Fig. \ref{fig:speed} shows that \texttt{qrpca} consistently perform faster than their related counterparts for datasets with more than 1,000 dimensions, with a top performance at least 15$\times$ faster for matrices with 10,000 dimensions and {\tt GPU} enabled. \texttt{sklearn} shows competitive performance against the python \texttt{qrpca} for high-dimensional data, while \texttt{prcomp} performs well on moderately large datasets, and the main advantage of using  \texttt{qrpca} comes from the {\tt GPU} on these cases. 
Overall, \texttt{qrpca} can be particularly beneficial for analyzing spectral data-cubes and hyperspectral images, and the next section shows one practical application in analyzing Astronomical spectra.

\subsection{MaNGA data}

Here we show one astronomical application using the IFU spectral data from the Mapping Nearby Galaxies at  Apache Point Observatory \citep[MaNGA;][]{bundy2015} survey. MaNGA is a program of the Sloan Digital Sky Survey IV  \citep[SDSS-IV;][]{blanton2017},  and also is the largest IFU survey of nearby galaxies to date. MaNGA obtains integrated field spectroscopy of galaxies using custom-designed fiber bundles, where the buffered fibers have a core diameter of 2 arcsec. Depending on galaxies' size, MaNGA uses different IFUs, where the IFU size range is from 19 to 127 fibers, and the corresponding field of view is from 12 to 32 arcsec in diameter. The wavelength coverage of MaNGA spectra is 3622–10354 $\AA$, and the resolution is about R $\sim2000$. For each galaxy target, MaNGA takes three dithered exposures. By stacking the spectra from dithered exposures, MaNGA builds a data cube ($N_X*N_Y*N_{wave}$) for each target galaxy, where the spatial pixel scale is designed as 0.5 arcsec per pixel \citep{Law_2016}. For the largest MaNGA IFU, the data cube has $N_X*N_Y=74*74$ spaxels. For the wavelength channel, the MaNGA data cube has $N_{wave}=4573$ for logarithmically sampled data and $N_{wave}=6732$ for the linear sampling \citep{Law_2016}.

\begin{figure*}[ht]
 \centering
\includegraphics[width=0.475\linewidth]{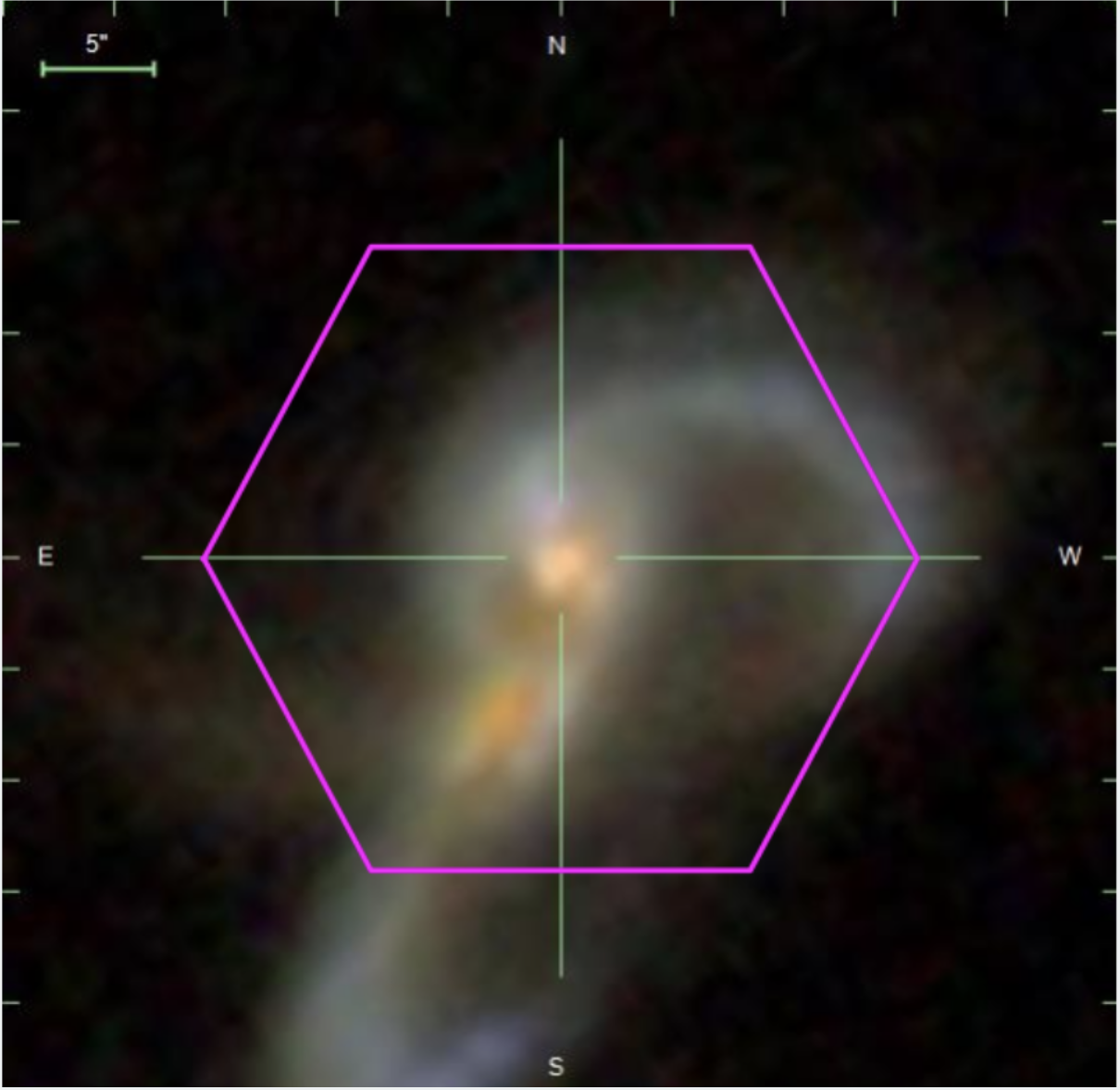}
\includegraphics[width=0.475\linewidth]{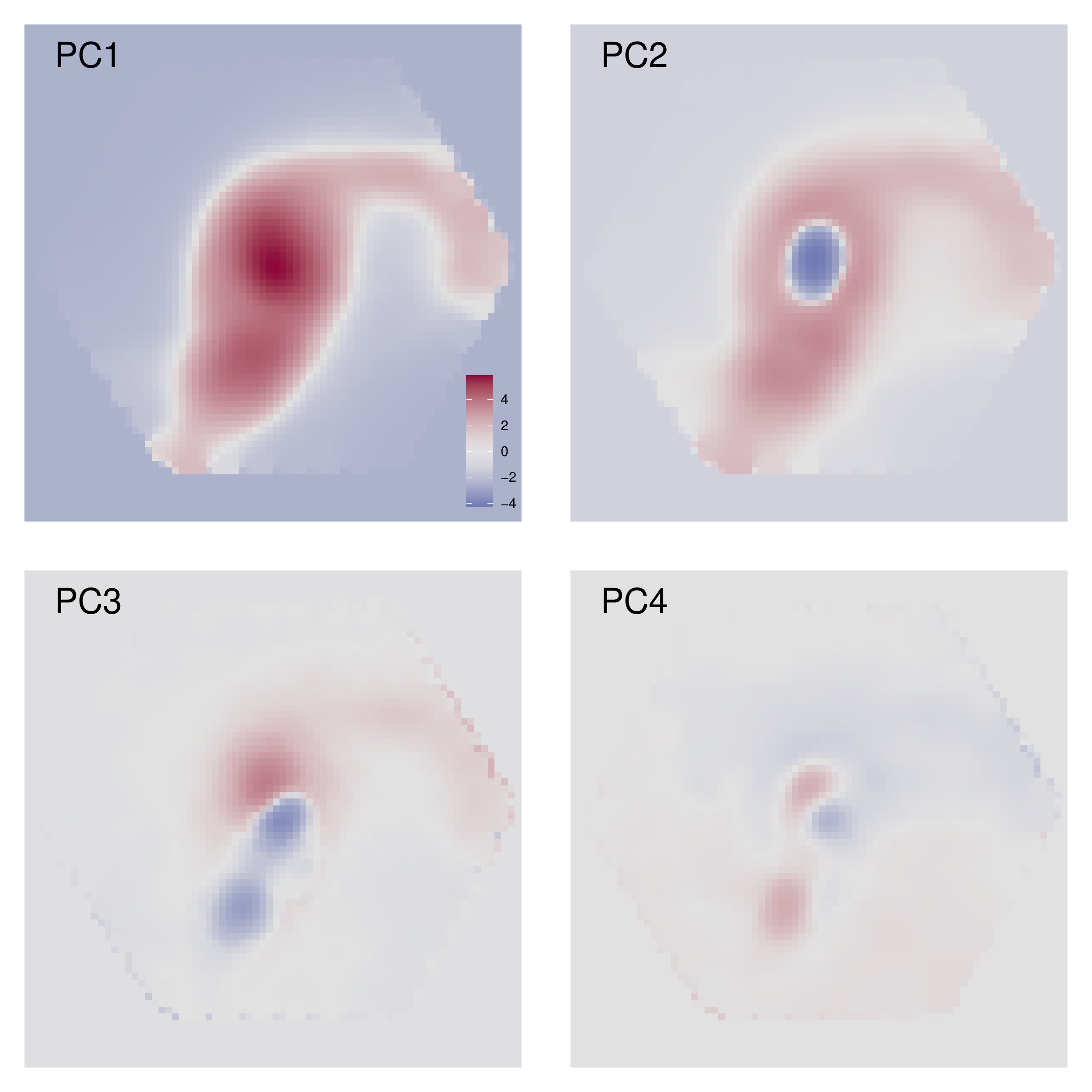}
\caption{Lef panel: The SDSS cutout of the galaxy Mrk 848 (MaNGA ID:12-193481) with a purple hexagon denoting the approximate spatial grasp of the IFU. Right panel display the first 4 eigenmaps, where different aspects of the merger structure can be seen.}
\label{fig:PCA_MaNGA}
\end{figure*}

We now show a simple task of computing the first eigenmaps of the MaNGA IFU data.  PCA decomposition of data-cubes is particularly interesting to disentangle uncorrelated physical phenomena in the galaxy. For example, it has been used to identify a broad line region of a previously unknown active galactic nucleus in the galaxy  NGC 4736 \citep{Steiner2009}. We showcase our approach by decomposing the IFU data of  the galaxy merger Mrk 848 (MaNGA ID:12-193481). Mrk 848 is a major merger with strong interaction between the two galaxies with estimated stellar masses $\log(M_{\star}/M_{\odot})$ = 10.44 and 10.30 \citep{Yang2007} at \emph{z} = 0.041. The MaNGA cube consists of an array of $74*74*4563$ array.

To decompose the data cube, we follow three basic steps. The first step involves transforming the tensor into a 5476$*$4563 matrix where each row represents one spectrum and a wavelength for each column. Then we apply a PCA transform to this matrix, which yields a matrix of a similar dimension, where each column now represents a PC. The final step is to transform back to the original format, and each PC will now represent an eigenmap. 
We show code snippets to read, process, and visualize the first eigenmaps with R and python versions of \texttt{qrpca} at \ref{sec:code}. The computation time with {\tt qrpca} is at least 2-3 times faster than standard implementations.

\autoref{fig:PCA_MaNGA} shows the first four PCs, where we can see different aspects of the merger structure. The first PC correlates with the overall merger structure, including the core and tail regions. The second PC isolates the core region of the central galaxy, while the third and fourth PCs discriminate the star-forming regions triggered by the merging process \citep{Yuan2018}. The galactic structures illuminated by this simple decomposition are broadly consistent with the different aged stellar populations revealed by detailed spectral energy distribution fitting  \citep{Yuan2018}, but further scrutiny of this object is beyond the scope of this work.

\section{Conclusions}
\label{sec:conclusion}
PCA is an essential tool for multivariate data analysis. However, its standard implementation does not scale well for high-dimensional datasets. In this paper, we present \texttt{qrpca}, a package for fast PCA computation based on QR decomposition. The code enables  {\tt GPU} acceleration when available. We showcase experiments on both simulated and real datasets of varying dimensions. Experimental results show that our package can perform more than 10 $\times$ faster than conventional approaches, depending on the matrix dimensions. 

\texttt{qrpca} is written in both {\tt R} and {\tt Python} and is freely available at GitHub\footnote{ \url{https://github.com/RafaelSdeSouza/qrpca}}, Zenodo\footnote{\url{https://doi.org/10.5281/zenodo.6556360}} and listed in the  
Python Package Index\footnote{\url{https://pypi.org/project/qrpca/}}.

\section*{Acknowledgments}
\label{sec:acknowledgements}
We thank Ana L. Chies-Santos for her insightful suggestions while preparing this manuscript. We thank Emille E. O. Ishida for the final revision of this manuscript. 
This work was supported by the National Natural Science Foundation of China (No. 1201101284, 12073059), the National Key R$\&$D Program of China (No. 2019YFA0405501), and the China Manned Space Project (No. CMS-CSST-2021-A04).

\appendix
\section{Code Snippets}
\label{sec:code}

Here, we show how to perform a PCA on MaNGA decomposition using the \texttt{qrpca} package. The IFU data used in our example is available at \url{https://data.sdss.org/sas/dr17/manga/spectro/redux/v3_1_1/7443/stack/manga-7443-12703-LOGCUBE.fits.gz}

\subsection*{R code for \texttt{qrpca} computation on MaNGA}
\begin{lstlisting}
require(qrpca);require(reticulate)
require(FITSio);require(ggplot2);
require(dplyr);require(reshape2)
cube <- "manga-7443-12703-LOGCUBE.fits"
df <- readFITS(cube)
n_row <- dim(df$imDat)[1]
n_col <- dim(df$imDat)[2]
n_wave <- dim(df$imDat)[3]
data.2D  <- array_reshape(df$imDat,
c(n_row*n_col,n_wave),order = c("F"))
pca  <- qrpca(data.2D)
# Function to extract the k-th eigenmap
eigenmap <- function(pcobj, k = 1){
  x <- as.matrix(pcobj$x)
  out <- matrix(x[,k],nrow=n_row,ncol=n_col)
  out
}

map1 <- eigenmap(pca) %>% melt()

ggplot(map1,aes(x=Var1,y=Var2,z=value)) +
  geom_raster(aes(fill=value)) +
  scale_fill_viridis_c(option="C") +
  theme(legend.position = "none") 
\end{lstlisting}

\subsection*{Python code for \texttt{qrpca} computation on MaNGA}

\begin{lstlisting}
from astropy.io import fits
from astropy.wcs import WCS
import torch
import numpy as np
from qrpca.decomposition import qrpca
import matplotlib.pyplot as plt

data = fits.open("manga-7443-12703-LOGCUBE.fits")
dat = data[1].data.transpose(1,2,0)
wcs = WCS(data[1].header)
da = dat.reshape(-1,dat.shape[-1]).astype(np.float32)
device = torch.device("cuda:0" if torch.cuda.is_available() else "cpu")
pca = qrpca(n_component_ratio=1,device=device)
map1 = pca.fit_transform(da)
ax = plt.subplot(projection=wcs[0,:,:])
ax = plt.gca()
lon = ax.coords['ra']
lat = ax.coords['dec']
plt.imshow(map1.reshape(74,74))
plt.xlabel("RA [deg]")
plt.ylabel("DEC [deg]")     
plt.show()
\end{lstlisting}

\bibliography{ref}

\end{document}